\documentclass[aps, pra,twocolumn,showpacs,amsmath,amssymb,longbibliography, 10pt]{revtex4-2}

\usepackage{tgtermes}
\usepackage[utf8]{inputenc}
\usepackage{bm}
\usepackage{color}
\usepackage{makeidx}
\usepackage{amsfonts}
\usepackage{amssymb}
\usepackage{amsthm}
\usepackage{graphicx}
\usepackage{epsfig}
\usepackage{subfigure}
\usepackage{amsmath,amssymb}
\usepackage[colorlinks=true,citecolor=blue,urlcolor=blue,linkcolor=blue]{hyperref}
\usepackage{xcolor}

\newcommand{\bea}{\begin{eqnarray}}
\newcommand{\ea}{\end{eqnarray}}
\newcommand{\eea}{\end{eqnarray}}

\newcommand{\sumint}[1]

\begin{document}
	
\title{Geometry-driven splitting dynamics of a triply quantized vortex in a ring-shaped condensate}
	
\author{Sixun Jia}
\author{Xin Wang}
\author{Xiaofeng Wu}
\author{Shuhang Wang}
\author{Bo Zhang}
\author{Bo Xiong}
\email{boxiong@whut.edu.cn}
\affiliation{Department of Physics, Wuhan University of Technology, Wuhan 430070, China}

\date{\today}

\begin{abstract}
We study the splitting dynamics of a triply quantized vortex (TQV) confined in a ring-shaped Bose-Einstein condensate under a weakly elliptical harmonic trap. Using full 3D simulations in cylindrical coordinates, combined with a semi-analytical energy analysis, we show that the vortex preferentially splits along the long axis of the trap, a direction that minimizes the kinetic-energy cost relative to the initial TQV state. Systematic parameter scans reveal that initial quantum fluctuations increase the splitting time and suppress the transient three-core pattern observed in noise-free simulations, whereas stronger nonlinear interactions accelerate the splitting. When the trap is nearly isotropic, the unstable Bogoliubov modes are dominated by both azimuthal quantum number $l_q=3$ and $l_q=2$; this leads to a dynamical sequence where three daughter vortices first form a triangular arrangement, later evolving into a linear chain. For stronger anisotropy, geometric coupling selectively enhances the $l_q=2$ mode, making it the sole dominant channel and resulting directly in linear vortex alignment---a clear signature of geometry-induced mode competition explained through combined energy-based and Bogoliubov stability analysis. Our results provide a quantitative picture of how trap geometry can steer the instability pathway, splitting time, and final pattern of a multiply quantized vortex, offering a route toward geometry-controlled vortex engineering.
\end{abstract}

\maketitle
\section{Introduction}
Quantized vortices, as topological defects carrying discrete units of circulation, are fundamental excitations in superfluids and Bose-Einstein condensates (BECs) \cite{FetterSvidzinsky2001,RevModPhys.81.647,Bewley2006,Anderson2010,PhysRevLett.90.140402}. Their creation, dynamical evolution, and mutual interactions have been a central theme in quantum gas research, offering a pristine and highly controllable platform for exploring superfluid hydrodynamics, topological matter, and out-of-equilibrium quantum dynamics \cite{PhysRevA.96.023617,PhysRevX.8.041056,8cdt-n3l2,Shen2019}. In these systems, singly quantized vortices (with winding number $S=1$) are typically stable topological excitations. In contrast, multiply quantized vortices ($S>1$) are intrinsically metastable: their energy scales approximately with $S^2$, making a configuration of $S$ well-separated singly quantized vortices energetically favorable over a solitary multiply charged vortex, under the constraint of total angular momentum conservation \cite{Pethick:2002a}. Consequently, a vortex with $S>1$ is generically prone to decay via dynamical instability into an array of $S$ singly quantized vortices. This splitting process is not a trivial disintegration but a rich, geometry-dependent dynamical reorganization of the superfluid, governed by the interplay of mean-field interactions, quantization conditions, and, crucially, the symmetry and shape of the external confinement. 

The stability of a multiply quantized vortex is governed by the spectrum of its collective Bogoliubov excitations \cite{Kuopanportti:2010a,Simula:2002a,Isoshima:2007a,PhysRevA.74.063619,KOBAYASHI20092359}. In a perfectly axisymmetric (isotropic) trap, the dominant decay channel for an $S$-charged vortex is typically triggered by the exponential growth of perturbations with a specific azimuthal symmetry, often leading to symmetric vortex arrays. These unstable modes are identified by complex eigenfrequencies in the linearized Bogoliubov--de Gennes (BdG) spectrum.
This idealized instability landscape is fundamentally reshaped by broken rotational symmetry. An elliptical deformation of the trapping potential couples different angular momentum channels, lifts mode degeneracies, and can selectively amplify particular unstable directions \cite{PhysRevA.95.043638,PhysRevA.81.023603,PhysRevA.77.053612,PhysRevA.100.033615,KOBAYASHI20092359}. Consequently, the trap geometry emerges as a decisive control parameter, capable of steering the instability to produce distinct, anisotropic vortex configurations. The initial trigger---whether an intentional seed or inherent fluctuations---then activates the instability along a direction set by the interplay of the perturbation's phase and the anisotropic confinement.

Despite this understanding, a systematic and predictive framework for how controlled trap ellipticity competes with or directs perturbation-seeded dynamics to dictate the splitting pathway of a specific higher-charge vortex remains incomplete. Studies have extensively mapped the decay of doubly quantized ($S=2$) vortices \cite{Xiong_2020}. In contrast, the $S=3$ case presents a richer and less charted territory. A central, unresolved question is: under what conditions does the decay of an $S=3$ vortex yield a symmetric triangular arrangement of three daughter vortices versus an anisotropic linear chain? This distinction is not merely morphological; it is crucial for applications in quantum information processing and analog gravity, where precise vortex geometries are required \cite{Wright2013,PhysRevLett.131.221602}. The current inability to reliably predict the final pattern from the initial trap shape and perturbation level motivates the present work. We therefore aim to establish a geometry-driven control mechanism for the splitting dynamics of a TQV, quantifying how trap anisotropy and fluctuation seeding selectively determine the instability pathway, its timescale, and the resulting vortex topology.

In this work, we investigate the splitting dynamics of a TQV with charge $S=3$ within a ring-shaped BEC, where the trap ellipticity serves as a precise control parameter. Our study combines full 3D Gross-Pitaevskii (GP) simulations---efficiently performed in cylindrical coordinates to capture the ring geometry---with analytical tools to establish a predictive framework for the instability. We show that even weak trap ellipticity robustly steers the TQV to split along the soft (long) axis, a preference explained by a semi-analytical energy comparison that reveals the kinetic-energy cost is minimized for this orientation. By systematically varying the anisotropy parameter $\eta$ and the interaction strength, we map the dependence of the splitting time $t_s$ and demonstrate that added quantum fluctuations delay the splitting, while stronger interactions accelerate it. Crucially, near the isotropic limit, the dynamics display a transient triangular vortex arrangement that subsequently evolves into a linear chain, a sequence directly linked to the presence of both $l_q=3$ and $l_q=2$ unstable modes in the Bogoliubov spectrum. As the anisotropy increases, the $l_q=2$ mode becomes dominant, leading directly to linear splitting---a clear manifestation of geometry-driven mode competition. Our results provide a comprehensive picture of how external confinement geometry can be used to  
direct the decay of a multiply quantized vortex, offering quantitative predictions for the splitting timescale and the resulting topological configuration. This work not only clarifies the instability mechanisms of high-charge vortices in anisotropic superfluids but also establishes practical  
guidelines for geometry-based control of vortex matter in quantum-gas experiments.

The remainder of this article is organized as follows. Section\,\ref{sec:methods} details our theoretical and numerical framework. We first introduce the initial state, i.e., a ring-shaped condensate carrying a TQV, and discuss its experimental preparation. We then present the 3D GP equation governing the dynamics, along with the cylindrical-coordinate numerical scheme used for its efficient solution. Finally, we outline the implementation of quantum fluctuations via the truncated-Wigner approximation. In section\,\ref{sec:results} we present our central results. We begin with the real-time splitting dynamics of the vortex, followed by a systematic analysis of how trap anisotropy, quantum noise, and interaction strength influence the splitting time. We then introduce a semi-analytical energy-based model that explains the preferential splitting axis, and conclude with a Bogoliubov stability analysis that links the observed splitting patterns to the underlying unstable excitation modes. Section\,\ref{C_O} summarizes our findings and discusses their implications.

\section{Model and methods}
\label{sec:methods}

We study the splitting dynamics of a TQV (charge $S=3$) created within a ring-shaped BEC and subsequently released into a weakly anisotropic (elliptic) harmonic trap. We quantify how (i) the trap ellipticity and (ii) fluctuation seeding modify the instability pathway and the characteristic splitting time. In this section, we detail the preparation of the initial ring-shaped condensate containing the vortex, the GP equation used to simulate its time evolution, and the numerical methods employed in our simulations.

\subsection{Initial ring BEC with $S=3$ circulation}
\label{subsec:initial}

The initial state in our simulation is a ring-shaped BEC containing a TQV along its symmetry axis. In cylindrical coordinates $(r,\phi,z)$, its wave
function is modeled as
\begin{equation} \label{wave1}
		\Psi (r, \phi, z) = \left\{ \begin{array}{cc} A \sqrt{ \mu - \frac{1}{2} m \omega_{r_0}^{2} r^{2}} f(r, z)  e^{i 3\phi}  &  \mu \geq \frac{1}{2} m \omega_{r_0}^{2} r^{2} \\
                                  0     &  {\rm otherwise}
                                     \end{array} \right.
\end{equation}
where $A$ is a normalization constant, $\mu$ is the chemical potential, $m$ is the atomic mass, and
$\omega_{r_0}$ is the radial trapping frequency of the underlying harmonic confinement. The
distribution $f(r,z)$ specifies the toroidal shape and axial profile:
\begin{equation}
\label{eq:distribution}
f(r,z)=\left[1-\exp\!\left(-a\frac{r^{2}}{2d^{2}}e^{-\frac{r^{2}}{2d^{2}}}\right)\right]
\exp\!\left(-\frac{z^{2}}{2\sigma_{z}^{2}}\right).
\end{equation}
Here, $\sigma_{z}=\sqrt{\hbar/(m\omega_{z})}$ is the axial Gaussian width corresponding to a trap
frequency $\omega_{z}$. The parameter $a$ tunes the size of the central density hole, effectively
matching the experimental profile shaped by a repulsive optical potential.

Triply quantized circulation can be prepared in an annular (toroidal) condensate, where the phase winding is
topologically protected. Experimentally, a toroidal trap can be realized by adding a blue-detuned ``plug''
beam that creates a repulsive barrier at the center of an otherwise harmonic magnetic trap \cite{49,50},
leading to an effective potential of the form
\begin{equation}
\label{eq:toroidal_trap}
V_{0}(r,z)=\frac{1}{2}m\omega_{r_0}^{2}r^{2}+\frac{1}{2}m\omega_{z}^{2}z^{2}
+V_{0}\exp\!\left[-\frac{2r^{2}}{d_{0}^{2}}\right].
\end{equation}
When $\mu<V_{0}$, a hole forms in the condensate center, yielding the ring geometry. This setup allows smooth
transition to a pure harmonic trap by reducing the laser power. The triply quantized phase can be realized
experimentally by phase imprinting technique or rotating the toroidal trap \cite{49,50,Kumakura:2006a,Leanhardt:2002a,Kuopanportti2010}. To access the subsequent splitting dynamics in a simply connected trap, one may adiabatically reduce the plug intensity,
thereby transforming the toroidal confinement into an (approximately) harmonic trap while preserving the
circulation. The ellipticity used in this work can then be tuned by adjusting the transverse frequencies
from $\omega_x=\omega_y$ to $\omega_x\neq\omega_y$.

The parameters for the initial ring-shaped condensate are selected within typical experimental
ranges for $^{87}\mathrm{Rb}$ atoms. We consider a total atom number $N = 10^{5}$. The trapping
frequencies that define the initial state via Eqs.~(1) and (2) are set to
$\omega_{r_0} = 2\pi \times 47\,\mathrm{Hz}$ and $\omega_{z} = 2\pi \times 238\,\mathrm{Hz}$,
yielding a condensate with an outer radius of $\sim 12\,\mu\mathrm{m}$ and an inner hole of
$\sim 6.5\,\mu\mathrm{m}$, consistent with the characteristic scale of ring BEC experiments \footnote{Private communication}. A typical density profile of the ring-shaped condensate is given in Fig.\,\ref{fig:3d}\,(a).

\subsection{Theoretical framework}

To study the subsequent dynamics, the prepared ring-shaped condensate is released into a
harmonic trap whose radial symmetry can be controllably broken. The time evolution of the
condensate wave function $\Psi(\mathbf{r},t)$ is governed by the 3D GP equation,
\begin{equation}
\label{eq:GPE}
i\hbar \frac{\partial \Psi}{\partial t}
=
\left[
-\frac{\hbar^{2}}{2m}\nabla^{2} + V(\mathbf{r}) + g|\Psi|^{2}
\right]\Psi,
\end{equation}
with an anisotropic external potential
\begin{equation}
\label{eq:anisotropic_potential}
V(\mathbf{r})
=
\frac{1}{2}m\omega_{r}^{2} r^{2}\left[1-\eta\sin^{2}\phi\right]
+\frac{1}{2}m\omega_{z}^{2} z^{2}.
\end{equation}
Here, $\omega_{r}=2\pi\times 79~\mathrm{Hz}$ sets the overall radial scale. The anisotropy
parameter $\eta$ is defined as
\begin{equation}
\eta=\frac{\omega_x^{2}-\omega_y^{2}}{\omega_x^{2}},
\end{equation}
where $\omega_x$ and $\omega_y$ are the trap frequencies along the $x$ and $y$ axes. In our
study $0\le \eta \le 1$; $\eta=0$ corresponds to an isotropic radial trap, while $\eta=1$
represents the limiting case of a potential that is quadratic along $x$ and uniform along $y$.
The interatomic interaction strength is
\begin{equation}
g=\frac{4\pi\hbar^{2}a_s}{m},
\end{equation}
where the $s$-wave scattering length for $^{87}\mathrm{Rb}$ is $a_s=5.4~\mathrm{nm}$ and the
atomic mass is $m=1.44\times 10^{-25}~\mathrm{kg}$. We vary $g$ from $0.1g_0$ to $50g_0$
(with $g_0$ being a reference coupling) to explore different interaction regimes.

To account for the effect of quantum fluctuations, we employ the truncated Wigner
approximation (TWA). Within this phase-space method, quantum (or thermal) fluctuations of
the Bose field are incorporated by stochastic sampling of initial conditions \cite{Steel1998,Castin:2002wkl,KOBAYASHI20092359}. The subsequent dynamics for each stochastic trajectory is then governed by the deterministic GP equation [Eq.~\eqref{eq:GPE}]. Concretely, each trajectory is initialized as
\begin{equation}
\label{eq:TWA_initial}
\Psi(\mathbf{r},0)=\Psi_{0}(\mathbf{r})+\xi(\mathbf{r}),
\end{equation}
where $\Psi_{0}$ is the mean-field ground state (our initial ring-shaped vortex state), and
$\xi(\mathbf{r})$ is a complex Gaussian random noise field with zero mean,
$\langle\xi(\mathbf{r})\rangle=\langle\xi^{*}(\mathbf{r})\rangle=0$ , and correlations corresponding to half a quantum per mode in the Wigner representation.

\subsection{Numerical methods}

\subsubsection{Cylindrical coordinate solver}
Simulating the dynamics of a giant vortex in a ring-shaped BEC presents significant numerical
challenges. The region near $r=0$, where vortex lines may merge, involves high momenta and is highly
sensitive to trap anisotropy. In Cartesian coordinates, accurately resolving this requires an
impractically fine grid to mitigate discretization errors associated with the square lattice.

Leveraging the initial cylindrical symmetry of the problem, we instead solve the GP equation in
cylindrical coordinates $(r,\phi,z)$. This approach enables efficient and accurate propagation on a grid of $320\times 200\times 100$ points in $(r,\phi,z)$ using a split-step Crank-Nicolson algorithm parallelized via the Message Passing Interface (MPI).
The results are subsequently transformed to Cartesian coordinates for analysis and visualization. For
our problem, this cylindrical scheme reduces the computational time by at least a factor of five
compared to an equivalent Cartesian simulation while delivering stable results. We note that advanced
methods like discrete exterior calculus on tetrahedral tilings have been proposed to further reduce grid-related artifacts for giant vortices \cite{Rabina:2018a}; our method provides a highly efficient
and reliable alternative for the present geometry.

\subsubsection{Implementation of quantum noise}
The TWA initial condition [Eq.~\eqref{eq:TWA_initial}] is implemented numericallyby adding the complex noise field $\tilde{\xi}(\mathbf{k})$ in the momentum-space low-energy subspace, ensuring a physical ultraviolet cutoff. The procedure is as follows:
\begin{enumerate}
\item The initial mean-field wave function $\Psi_{0}$, defined in cylindrical coordinates
  $(r,\phi,z)$, is first interpolated onto a 3D Cartesian grid $(x,y,z)$.
\item A 3D Fourier transform yields the momentum-space representation $\tilde{\Psi}_{0}(\mathbf{k})$.
\item The complex noise field $\xi(\mathbf{r})$ is constructed in momentum space. We expand it
  in plane-wave modes up to a chosen energy cutoff $E_{\mathrm{cut}}$:
  \begin{equation}
    \xi(\mathbf{r})=\sum_{|\mathbf{k}|\le k_{\mathrm{cut}}}\eta_{\mathbf{k}}\,e^{i\mathbf{k}\cdot\mathbf{r}},
  \end{equation}
  where the complex amplitudes $\eta_{\mathbf{k}}$ are independent Gaussian random variables
  satisfying
  \begin{equation}
    \langle \eta_{\mathbf{k}} \rangle = 0,
    \qquad
    \langle \eta_{\mathbf{k}}\eta_{\mathbf{k}'} \rangle = 0,
    \qquad
    \langle \eta_{\mathbf{k}}\eta_{\mathbf{k}'}^{*} \rangle = \frac{1}{2}\delta_{\mathbf{k}\mathbf{k}'}.
  \end{equation}
\item The noisy wave function $\tilde{\Psi}_{0}(\mathbf{k})+\tilde{\xi}(\mathbf{k})$ is transformed
  back to real-space Cartesian coordinates.
\item Finally, the state is re-interpolated onto our primary cylindrical coordinate grid $(r,\phi,z)$
  for efficient deterministic propagation using the split-step Crank--Nicolson algorithm parallelized via the MPI.
\end{enumerate}

\begin{figure*}[t]
  \centering
  \includegraphics[width=\textwidth]{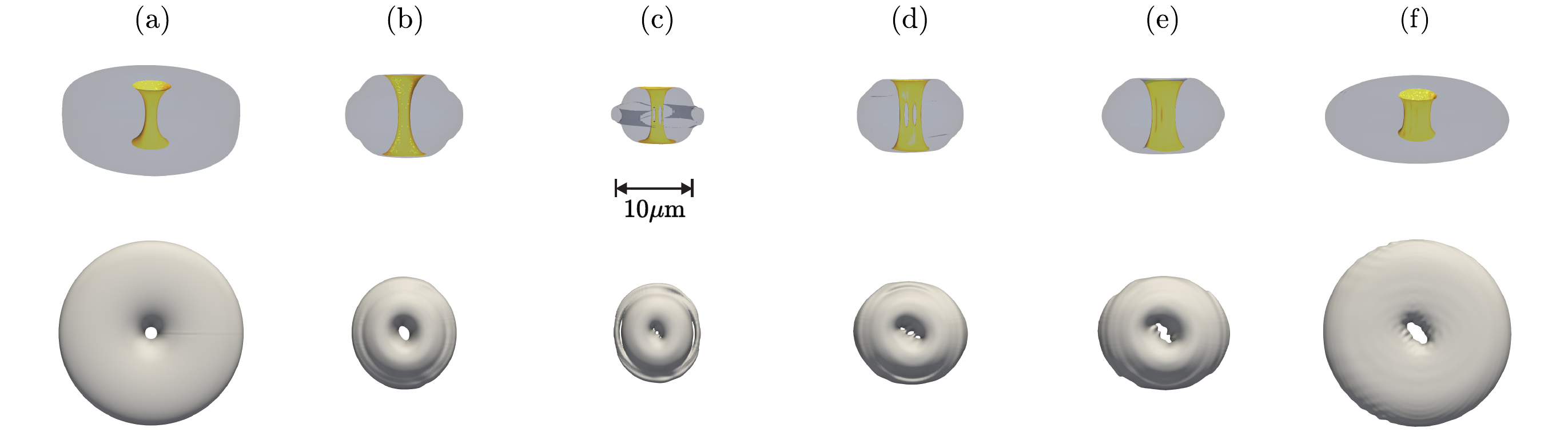}
  \caption{(Color online) Representative 3D splitting and subsequent fusion of an initially imprinted
$S=3$ vortex in an elliptic trap with anisotropy $\eta=0.1$.
Top row: side-view renderings of the condensate isodensity surface $\rho(\mathbf r)=|\psi(\mathbf r)|^{2}$ at
$\rho=0.3\%\,\rho_{\max}$. The same surface is shown with reduced opacity in the outer region (gray), while a cropped
region around the vortex core is highlighted (yellow) to emphasize the fragmentation and reconnection of the multiply
quantized core.
Bottom row: corresponding top-view density snapshots used to classify the splitting stage by the number of resolved vortex holes. 
Times are (a) $t=0~\mathrm{ms}$, (b) $t=0.875~\mathrm{ms}$, (c) $t=0.981~\mathrm{ms}$, (d) $t=1.512~\mathrm{ms}$,
(e) $t=1.592~\mathrm{ms}$, and (f) $t=2.122~\mathrm{ms}$.}
\label{fig:3d}
\end{figure*}

\section{Results and discussion}\label{sec:results}

\subsection{Representative dynamics}

We begin by presenting the characteristic dynamics of an initially ring-shaped BEC with $S=3$
circulation released into an elliptic trap with anisotropy parameter $\eta=0.1$.
Fig.~\ref{fig:3d} displays the full 3D evolution of the condensate density, rendered via an
isosurface fixed at $0.3\%$ of the peak density.
The evolution proceeds through several distinct stages. Initially rotationally symmetric [see (a)],
the cloud is deformed elliptically by the trap [see (b)], while the vortex core remains intact---visible
as a single density hole in the top view and an unbroken dumbbell-shaped depletion in the side view. As the elliptic deformation develops, the instability sets in and the triply quantized core rapidly breaks up.
Subsequently, the TQV undergoes splitting: three separate vortex holes emerge
in the top view [see (c) and (d)] and the side-view dumbbell develops two clear ruptures, with the
separation between holes increasing over time.
Notably, this splitting process differs qualitatively from dynamics reported in cigar-shaped
condensates \cite{Mateo:2006a,Huhtamaki:2006a}, where a pair of singly charged vortices emerging from a
doubly quantized vortex typically twist around one another along the longitudinal axis. In our
elongated ring-shaped geometry, such intertwining of the splitting vortex lines does not occur. This
suggests that the collective excitation spectrum of the ring-shaped condensate suppresses higher
unstable modes, leaving only the lowest-frequency instability active, which drives a rapid axial
($z$-direction) splitting.We emphasize that a finite anisotropy ($\eta>0$) is essential for
triggering this instability; in the isotropic limit ($\eta=0$), the TQV remains stable and does not
split into three singly quantized vortices, although the BEC itself undergoes breathing oscillations
(see Fig.\,\ref{fig:process_of_eta=0} in Appendix.\,\ref{app:TQV_stability_isotropic}). At later times, breathing motion of the cloud rapidly depletes atoms near the $z$-axis, creating a
local imbalance in the interaction forces on the vortices. This drives the three singly quantized
vortices to recombine into a single giant vortex [see (e) and (f)], completing a clear
splitting--recombination cycle. Throughout this work, we define the splitting time $t_s$ as the time
elapsed from the initial state to the first unambiguous appearance of three separated vortex cores.
This panel-resolved identification, based on tracking the vortex topology from the start, allows us
to objectively determine $t_s$ and distinguish the onset of splitting from later dynamical events.

\begin{figure}[t]
  \centering
  \includegraphics[width=\linewidth]{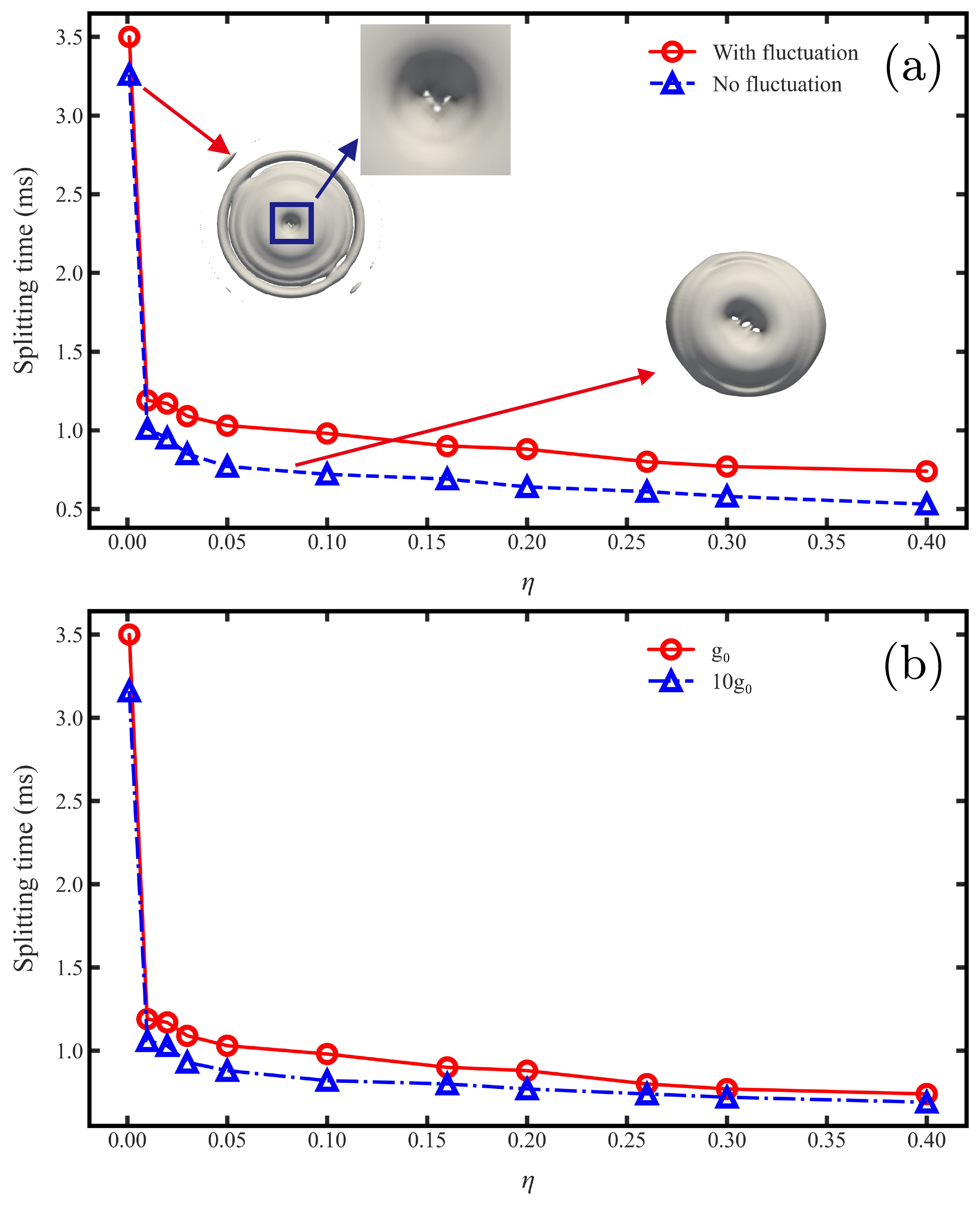} 
  \caption{(Color online) Splitting time of the TQV.
  (a) Splitting time $t_s$ as a function of trap anisotropy $\eta$ for a fixed interaction strength $g_0$.
  Blue squares: dynamics from the mean-field initial state (unseeded).
  Red circles: dynamics including initial quantum fluctuations via the truncated Wigner approximation (seeded).
  The schematics in the inset illustrate the characteristic splitting patterns for the near-isotropic limit
  ($\eta \to 0$, top left) and for finite anisotropy (bottom right).
  (b) Splitting time $t_s$ versus $\eta$ in the presence of quantum fluctuations for two interaction strengths, $g_0$ (red circles) and $10g_0$ (blue triangles).}
  \label{fig:ts_eta_g_seed}
\end{figure}

\subsection{Splitting time versus trap anisotropy}
We first examine the influence of quantum fluctuations on the splitting dynamics. Fig.\,\ref{fig:ts_eta_g_seed} (a)
reveals a systematic delay in the splitting time $t_s$ when initial quantum fluctuations are
included via the TWA, compared to the purely mean-field (unseeded) evolution. For a fixed anisotropy $\eta$, the seeded simulations consistently yield a longer $t_s$. This delay can be attributed to the role of initial quantum noise. The added fluctuations generate quasiparticle excitations, particularly within and around the vortex core region. These excitations interact with the nascent vortex fragments, effectively repelling or
deforming them and thus obstructing the clean, coordinated separation required to establish
three distinct, stable cores. Consequently, the early-stage dynamics in the presence of noise
are characterized by intermittent events of partial splitting and subsequent reconnection,
rather than a direct transition to a persistent triple-vortex state. Since $t_s$ is defined as
the first moment a persistent three-core topology is established, this intermittent behavior naturally postpones its measured onset.
Our analysis indicates that the introduced quantum perturbations do not merely seed the
underlying instability; they can actively impede the coherent assembly of a well-defined,
separated triple-vortex configuration. In both cases, $t_s$ decreases sharply as $\eta$ increases from near-zero (the
quasi-isotropic limit), followed by a more gradual variation at larger $\eta$. This trend is
consistent with the picture that even a weak elliptic deformation efficiently mixes angular
harmonics and selects a preferred orientation for the instability of the $S=3$ core, thereby
enhancing its coupling to unstable splitting modes. The strongest reduction in $t_s$ occurs just
away from isotropy; at higher $\eta$, the decrease moderates, suggesting a crossover from a
symmetry-breaking--limited regime to one where the instability growth is governed primarily by
nonlinear and kinetic energy scales.

The distinct splitting patterns of the TQV are directly influenced by the trap
geometry. Fig.\,\ref{fig:ts_eta_g_seed} (a) contrasts the outcomes for two limiting cases. Near the isotropic
limit ($\eta \to 0$), the instability preferentially populates a mode that leads to a triangular
arrangement of the three resulting singly quantized vortices (top schematic). In contrast,
for finite anisotropy ($\eta>0$), the splitting follows the principal axes of the elliptical
trap, yielding a linear alignment of the vortex cores along the soft confinement direction (bottom
schematic). This clear geometric dependence---from a triangular to a linear configuration---motivates
the analysis in Sec.\,\ref{sec:results}, where we examine the energetic preference for the splitting direction.
Furthermore, the stability of these distinct patterns (the $l_q=3$ triangular mode versus the $l_q=2$
linear mode) will be analyzed within the Bogoliubov framework in Sec.\,\ref{C_O}.

We next examine the role of interatomic interaction strength. Fig.\,\ref{fig:ts_eta_g_seed} (b) shows that, under the same fluctuation-seeding protocol, a stronger nonlinear coupling (here, $10g_0$ versus a reference value) reduces $t_s$ across all $\eta$. This trend aligns with the expectation that enhanced
nonlinear mode coupling accelerates the growth of the splitting instability in a more strongly
interacting condensate. 

\begin{figure*}[t]
  \centering
  \includegraphics[width=\textwidth]{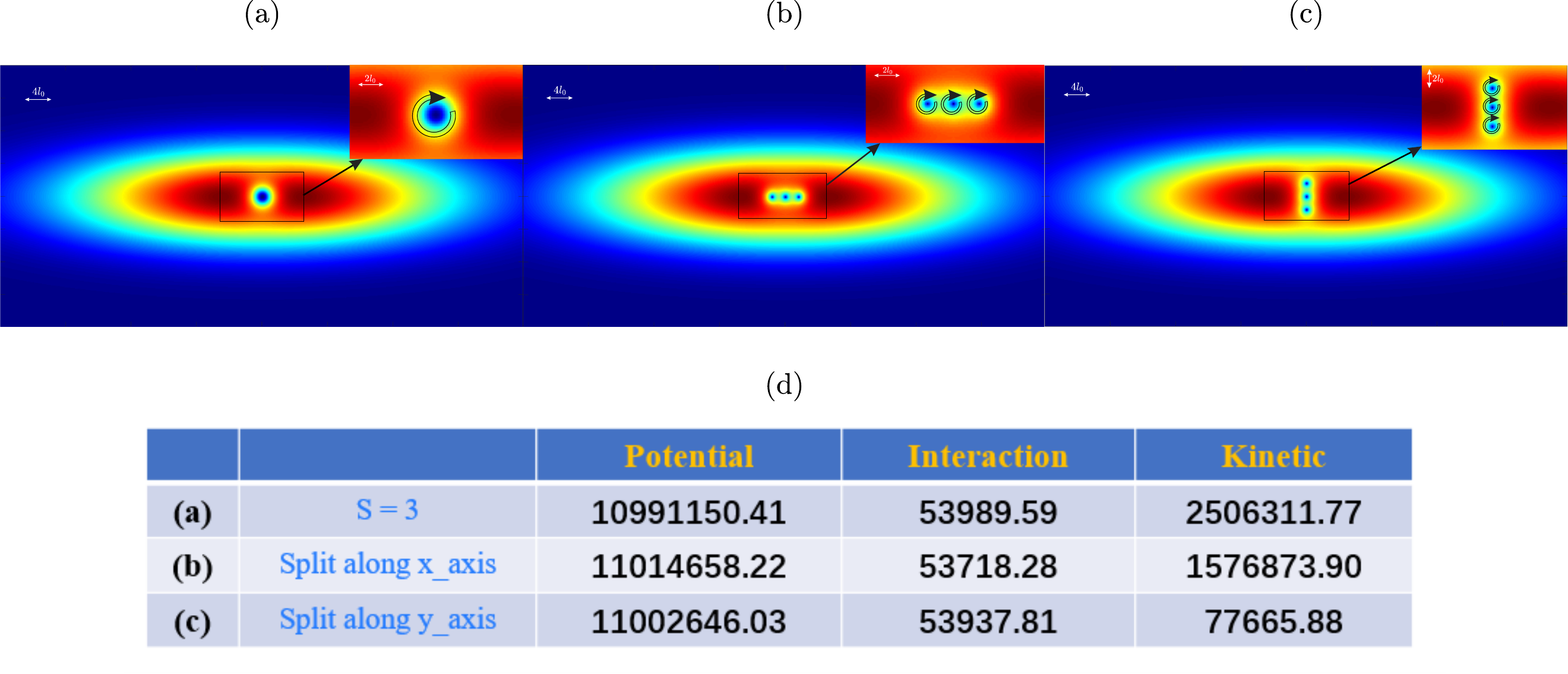}
  \caption{(Color online) Energetic comparison of candidate post-splitting configurations.
  Top: representative $xy$-plane density profiles for (left) an intact $S=3$ vortex, (middle) a fully split
  configuration aligned along the $x$ axis, and (right) a fully split configuration aligned along the
  $y$ axis. Insets show a magnified view of the vortex cores.
  Bottom: decomposition of the total energy into potential, interaction, and kinetic contributions for
  the three configurations (semi-analytical evaluation using the corresponding stationary density
  profiles). The potential and interaction energies are comparable across configurations, whereas the
  kinetic energy discriminates the splitting direction, favoring the $x$-axis split state for the parameters shown.}
  \label{fig:energy_axis}
\end{figure*}

\subsection{Energetic Preference for the Splitting Direction}
\label{sbus:enery}

To understand the preferred splitting direction observed in the dynamics, we analyze the energy
landscape of possible vortex configurations within an effective 2D model. We
consider the condensate field $\psi(x,y)$, which represents a transverse slice (or the result of
integrating out the $z$ coordinate under the Gaussian axial profile
$\exp\!\left(-z^{2}/2\sigma_{z}^{2}\right)$). In this reduction, the effective 2D interaction
coefficient is $g_{2D}=g/\sqrt{{2\pi}\,\sigma_{z}},\sigma_{z}=\sqrt{{\hbar}/{m\omega_{z}}}$.
All quantities are expressed in natural units of the harmonic oscillator: lengths in units of the
radial harmonic length $a_{\mathrm{ho}}=\sqrt{{\hbar}/{m\omega_{x}}}$, and energies in units of $\hbar\omega_{x}$.
To compare candidate post-splitting configurations on an equal footing, we construct simple trial
wave functions for (i) an intact $S=3$ vortex, and (ii) three fully separated singly quantized
vortices aligned either along the $x$ or $y$ axis. The overall density envelope is modeled by a
normalized Gaussian,
\begin{equation}
\label{eq:gauss2d}
G(x,y)=\frac{1}{2\pi \sigma_x \sigma_y}
\exp\!\left(-\frac{x^2}{2\sigma_x^2}-\frac{y^2}{2\sigma_y^2}\right),
\end{equation}
with dimensionless widths (in units of $a_{\mathrm{ho}}$) $\sigma_x=20$ and $\sigma_y=5$,
chosen to match the radial extent of the ring-shaped condensate in our simulations. The
vortex-core profile is approximated by the function
\begin{equation}
\label{eq:vortexcore_profile}
F(r)=\frac{r}{\sqrt{2+r^2}}, \qquad r=\sqrt{x^2+y^2},
\end{equation}
and we define the polar angle $\theta=\arg(x+iy)$ \cite{Pethick:2002a}. The intact TQV state is then modeled as
\begin{equation}
\label{eq:trial_S3}
\psi_{S=3}(x,y)=\mathcal{N}\,G(x,y)\,F(r)\,e^{i3\theta},
\end{equation}
where $\mathcal{N}$ is a normalization constant fixed to maintain the same 2D particle number
for all configurations.

To model a fully split state, we superpose three singly quantized vortices at prescribed core
positions. For a vortex centered at $\mathbf{r}_j=(x_j,y_j)$, we define
\begin{equation}
\label{eq:local_polar}
r_j=\sqrt{(x-x_j)^2+(y-y_j)^2}, \qquad
\theta_j=\arg\!\bigl((x-x_j)+i(y-y_j)\bigr).
\end{equation}
A singly quantized vortex factor is $F(r_j)e^{i\theta_j}$. Following the geometric splitting
patterns observed, the ``$x$-axis'' (long-axis) split state is constructed by placing the three
vortex cores at $x_j\in\{-d,0,+d\}$ and $y_j=0$. Similarly, the ``$y$-axis'' (short-axis) split
state places the cores at $y_j\in\{-d,0,+d\}$ and $x_j=0$. The trial wave function is
\begin{equation}
\label{eq:trial_splitx}
\psi_{x\text{-split}}(x,y)=\mathcal{N}\,G(x,y)\left[
\prod_{j=1}^{3} F(r_j)\,e^{i\theta_j}
\right]^{1/3},
\end{equation}
where the product runs over the three vortex positions. The dimensionless separation $d=2$
(in units of $a_{\mathrm{ho}}$) is chosen such that the resulting density patterns match the fully split configurations observed in the dynamical simulations. The energetic comparison is valid in the regime where the vortex-core separation $d$ is of the same order as the healing length. With the chosen value $d=2$, the vortex cores are well resolved while the separation remains much smaller than the characteristic density-variation length scale of the condensate, justifying a local energetic analysis.

The total 2D energy (in units of $\hbar\omega_r$) is decomposed as
\begin{equation}
E_{\mathrm{kin}}=\frac{1}{2}\iint |{\nabla\psi}|^{2}\,dx\,dy,
\qquad
\end{equation}
\begin{equation}
E_{\mathrm{pot}}=\frac{1}{2}\iint \left(\tilde{\omega}_x^{2}x^{2}+\tilde{\omega}_y^{2}y^{2}\right)|{\psi}|^{2}\,dx\,dy,
\end{equation}
\begin{equation}
E_{\mathrm{int}}=\frac{g'_{2\mathrm{D}}}{2}\iint |{\psi}|^{4}\,dx\,dy,
\end{equation}
with $\tilde{\omega}_x=\omega_x/\omega_r=1$, $\tilde{\omega}_y=\omega_y/\omega_r=\sqrt{1-\eta}$,
and $g'_{2\mathrm{D}}=g_{2\mathrm{D}}/(\hbar\omega_r a_{\mathrm{ho}}^{2})$ being the dimensionless
2D interaction strength. The total energy is
\begin{equation}
E_{\mathrm{total}}=E_{\mathrm{kin}}+E_{\mathrm{pot}}+E_{\mathrm{int}}.
\end{equation}

Figure\,\ref{fig:energy_axis} compares the energy decomposition for three configurations: the intact $S=3$ vortex
[(a)], the fully split state aligned along the $x$-axis [(b)], and the fully split
state aligned along the $y$-axis [(c)]. The corresponding energy contributions are shown
in Fig.\,\ref{fig:energy_axis} (d). While the potential and interaction energies are nearly identical among the three
configurations, the kinetic energy differs significantly. Crucially, the kinetic energy of the
$x$-axis split state is much closer to that of the intact vortex than the kinetic energy of
the $y$-axis split state.
This kinetic-energy proximity has direct dynamical implications. In a closed quantum system,
total energy is conserved. Therefore, a structural transformation from the initial intact vortex
to a final split state must approximately preserve the total energy. The configuration whose total
energy (dominated by the kinetic component in this case) is closest to the initial one is
energetically more accessible. The significant kinetic-energy difference between the intact vortex
and the $y$-axis split state would require a substantial compensating change in other energy forms
(potential or interaction) during the dynamical transition, which is not favored. In contrast, the
smaller kinetic-energy mismatch for the $x$-axis split state makes it the preferred pathway,
consistent with the observed linear alignment of vortices along the trap's soft direction.
This analysis clarifies that the selection of the splitting direction is primarily governed by the
kinetic-energy cost, which favors fragmentation along the direction that minimizes the deviation
from the initial kinetic energy.

\subsection{Symmetry Content of Unstable Modes and Mode Competition}

\begin{figure}[t]
  \centering
  \includegraphics[width=\linewidth]{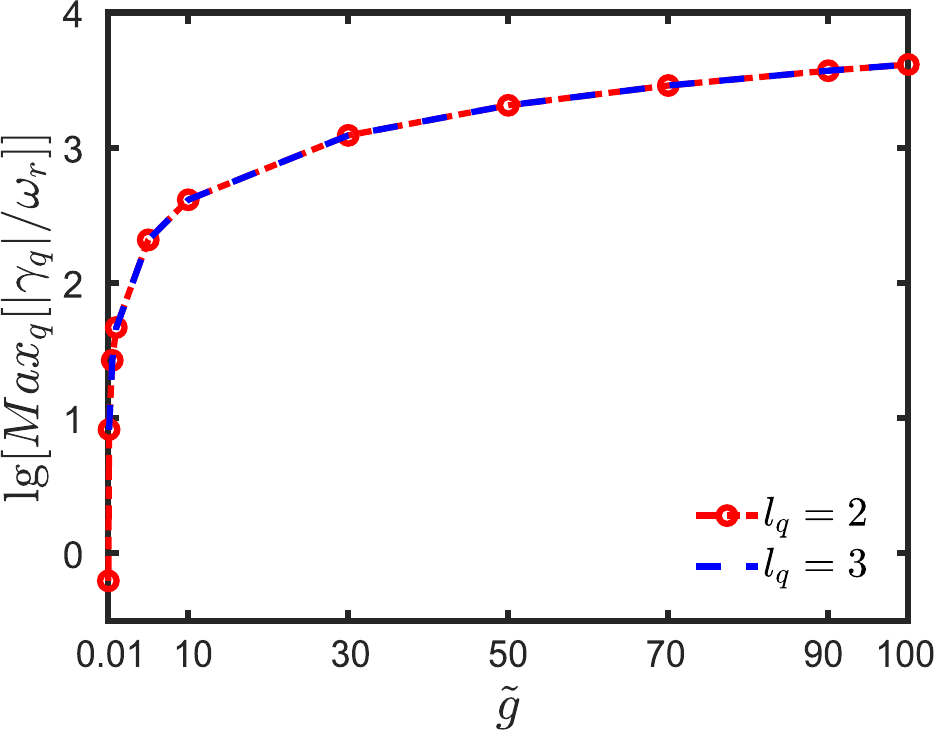}
  \caption{(Color online) Logarithmic scale of the maximum imaginary part (growth rate $\gamma$) of the unstable Bogoliubov eigenfrequencies as a function of the dimensionless interaction strength $\tilde{g}$ for the axisymmetric ($\eta=0$) ring condensate. The curves correspond to different azimuthal quantum numbers of the perturbation: $l_q=2$ (red dashed line with circles) and $l_q=3$ (blue dashed line).}
  \label{fig:Im}
\end{figure}

To identify the microscopic origin of the splitting patterns observed in our 3D simulations, we analyze the collective excitation spectrum of the initial $S=3$ vortex state. Given the strong axial confinement ($\omega_z \gg \omega_r$) and the fact that the vortex line is aligned with the $z$-axis, transverse (radial) instabilities dominate the splitting dynamics. We therefore consider an effective 2D description, obtained by integrating out the Gaussian axial profile, where the stationary vortex state is
\[
\psi_0(r,\phi)=f(r)e^{iS\phi}.
\]
Here, $f(r)$ corresponds to the radial density profile of the ring-shaped condensate, consistent with the initial condition used in the 3D simulations.

We examine small perturbations to this stationary state within the Bogoliubov--de Gennes (BdG) framework. The perturbed wave function is written as
\begin{equation}
\begin{split}
\psi(r,\phi,t)&=e^{-i\mu t}\Bigl[ 
f(r)e^{iS\phi} + u_q(r)e^{i(S+l_q)\phi-i\omega_q t} \\
&+ v_q^{*}(r)e^{i(S-l_q)\phi+i\omega_q t}
\Bigr],
\end{split}
\label{eq:bdg_ansatz}
\end{equation}
where $\mu$ is the chemical potential, and $(u_q,v_q)$ are the radial Bogoliubov amplitudes corresponding to an eigenmode with azimuthal quantum number $l_q$ and complex frequency $\omega_q$. These amplitudes satisfy the normalization condition
\begin{equation}
\int_0^{\infty} r\,dr \left[ |u_q(r)|^2 - |v_q(r)|^2 \right] = 1.
\end{equation}

Substituting Eq.~\eqref{eq:bdg_ansatz} into the GP equation and linearizing yields the radial BdG eigenvalue problem
\begin{equation}
\begin{pmatrix}
\mathcal{L}_{S+l_q} & g_{\mathrm{2D}} f^2(r) \\[4pt]
- g_{\mathrm{2D}} f^2(r) & -\mathcal{L}_{S-l_q}
\end{pmatrix}
\begin{pmatrix}
u_q(r) \\
v_q(r)
\end{pmatrix}
=
\hbar\omega_q
\begin{pmatrix}
u_q(r) \\
v_q(r)
\end{pmatrix},
\label{eq:bdg}
\end{equation}
where the single-particle operator $\mathcal{L}_m$ is defined as
\begin{equation}
\mathcal{L}_m
=
-\frac{\hbar^2}{2m}
\left(
\frac{d^2}{dr^2}
+\frac{1}{r}\frac{d}{dr}
-\frac{m^2}{r^2}
\right)
+V_{\perp}(r)-\mu+2g_{\mathrm{2D}}f^2(r).
\label{eq:Lk}
\end{equation}
This operator incorporates the kinetic energy (including the centrifugal barrier $\propto m^2/r^2$), the transverse confining potential $V_{\perp}(r)$, the chemical potential, and the Hartree mean-field correction. The effective 2D interaction strength is $g_{\mathrm{2D}}={g}/{\sqrt{2\pi}\,\sigma_z},$ as introduced in Sec.\,\ref{sbus:enery}. A dynamical instability is signaled by the appearance of complex eigenfrequencies $\omega_q=\omega_q^{(r)}+i\gamma_q$. The imaginary part $\gamma_q>0$ gives the growth rate of the corresponding mode; the larger $\gamma_q$, the faster the perturbation grows and drives the vortex toward splitting. The azimuthal index $l_q$ of the most unstable mode directly predicts the symmetry of the incipient splitting pattern: $l_q=3$ corresponds to a threefold (triangular) deformation, whereas $l_q=2$ leads to a quadrupolar (linear) deformation.

Figure\,\ref{fig:Im} shows that in the isotropic limit ($\eta=0$) and across a broad range of interaction strengths, the maximum imaginary parts (growth rates $\gamma$) of the Bogoliubov spectrum for modes with $l_q=3$ and $l_q=2$ are nearly identical and increase with the interaction strength. This demonstrates that the mean-field interaction enhances the growth rate of these unstable collective modes. The near-degeneracy of the $l_q=2$ and $l_q=3$ growth rates in the isotropic trap implies that both triangular ($l_q=3$) and linear ($l_q=2$) deformations are, in principle, accessible instability channels. This theoretical picture, however, appears to contrast with the full 3D simulation for $\eta=0$, in which the TQV remains stable and does not decay into three singly quantized vortices (Appendix \,\ref{app:TQV_stability_isotropic}). The apparent discrepancy is resolved by recognizing that the periodic radial breathing of the ring-shaped condensate—driven by the initial toroidal preparation and the harmonic confinement—dynamically suppresses the exponential growth of the unstable Bogoliubov modes, effectively stabilizing the vortex core on the timescale of the simulation. Introducing even a weak trap anisotropy ($\eta>0$) breaks this protection and allows the instabilities to manifest. As shown in Appendix \,\ref{app:TQVV} for $\eta=0.001$, the dynamics initially display a transient triangular vortex arrangement (driven by the $l_q=3$ mode) that subsequently evolves into a linear chain (as the $l_q=2$ mode becomes relevant). With increasing anisotropy, the geometric coupling selectively enhances the $l_q=2$ channel, making it the dominant instability and leading directly to linear splitting—a clear signature of geometry-driven mode competition.

\section{Conclusion and outlook}
\label{C_O}
We have investigated the splitting dynamics of a TQV ($S=3$) initially embedded in a ring-shaped BEC and released into an elliptical harmonic trap. By combining full 3D GP simulations in cylindrical coordinates with complementary analytical tools, we have resolved the complete splitting-recombination cycle and established clear criteria for defining the splitting time $t_s$. Our parameter scans reveal that the splitting time $t_s$ decreases with increasing trap anisotropy $\eta$, as the elliptical deformation enhances the coupling to unstable collective modes. Introducing initial quantum fluctuations via the truncated Wigner approximation systematically increases $t_s$, indicating that the added noise can interfere with the coherent formation of a stable three-vortex state. Notably, in noise-free simulations very close to the isotropic limit ($\eta \to 0$), the dynamics exhibit a transient triangular arrangement of the three daughter vortices before they evolve into a linear chain. In contrast, a stronger nonlinear interaction reduces $t_s$ by accelerating the growth of the dominant instability. The preferred splitting direction-along the trap's soft (long) axis is explained through a semi-analytical energy decomposition. While the potential and interaction energies remain nearly unchanged between candidate configurations, the kinetic energy cost is minimized for the linear alignment along this axis, providing a direct energetic rationale for the observed dynamics. A Bogoliubov-de Gennes stability analysis of the axisymmetric ($\eta = 0$) vortex state clarifies the origin of the distinct splitting patterns. The spectrum contains competing unstable modes with azimuthal quantum numbers $l_q=3$ and $l_q=2$. In the nearly isotropic regime, both modes are present, leading to the observed dynamical sequence from a triangular ($l=3$) to a linear ($l_q=2$) arrangement. As the trap anisotropy increases, geometric coupling selectively amplifies the $l_q=2$ mode, making it the dominant channel and resulting directly in linear splitting. This crossover, driven by the competition between symmetry content and trap geometry, is consistently explained by our combined energy and linear stability analysis.

In summary, our work provides a quantitative framework for understanding how external confinement geometry governs the instability pathway, timescale, and final topological configuration of a multiply quantized vortex. It demonstrates that trap anisotropy can act as a reliable control knob to steer the decay of a high-charge vortex toward a desired pattern. Future studies could extend this approach by quantitatively linking the extracted growth rates to full Bogoliubov spectra in anisotropic traps, and by incorporating more refined beyond-mean-field noise models to further elucidate the interplay between fluctuations and geometry in vortex matter engineering.

\section*{Acknowledgment}
We would like to thank Junhui Zheng for many useful discussions and suggestions in this work. This work is supported by the National Natural Science Foundation of China under Grants No.12075175 and No.12575028. 

\appendix

\begin{figure*}[t]
  \centering
  \includegraphics[width=\textwidth]{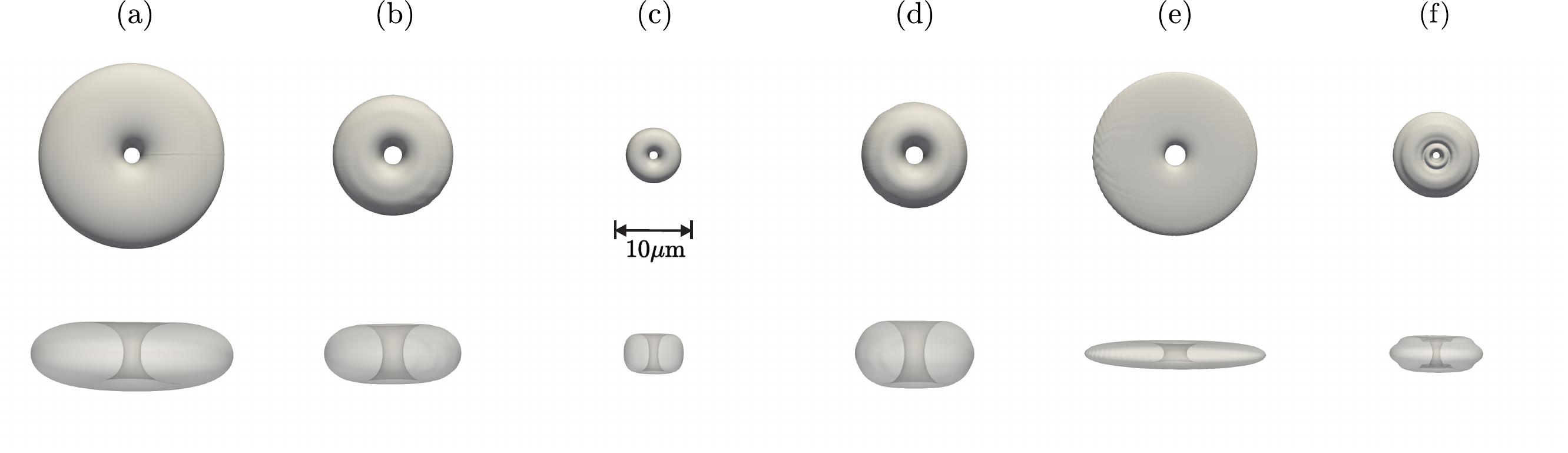}
  \caption{(Color online) Temporal density isosurface of a ring-shaped condensate with the circulation $S=3$ in a symmetric trap, i.e., $\eta=0$ for $1\%$ of the maximum density and $g=g_0$. Different panels correspond to different time : (a) $t=0~\mathrm{ms}$, (b) $t=0.663~\mathrm{ms}$, (c) $t=1.325~\mathrm{ms}$, (d) $t=1.59~\mathrm{ms}$, (e) $t=2.65~\mathrm{ms}$, and (f) $t=26.474~\mathrm{ms}$.}
  \label{fig:process_of_eta=0}
\end{figure*}

\section{Stability of the TQV for $\eta=0$}
\label{app:TQV_stability_isotropic}

This appendix provides supporting evidence that the TQV within a ring-shaped BEC can remain dynamically stable in a perfectly isotropic trap ($\eta=0$), in contrast to its well-known instability in a standard (non-ring) condensate. This stability forms the essential backdrop for the main text, where a finite trap anisotropy $\eta>0$ is introduced to controllably trigger the splitting instability.

We simulate the evolution of the initial ring-shaped condensate with $S=3$ circulation, prepared as described in Sec.\,\ref{sec:methods}, in an isotropic harmonic trap ($\eta=0$). Fig.\,\ref{fig:process_of_eta=0} illustrates this dynamics: the initial state with a central density hole [Fig.\,\ref{fig:process_of_eta=0}(a)] first undergoes a transient contraction, forming a well-defined, axisymmetric TQV at the center [Fig.\,\ref{fig:process_of_eta=0}(b)]. Subsequently, the condensate exhibits robust collective excitations—primarily a radial breathing mode and a quadrupole oscillation—manifested as periodic modulations in the outer density profile. Crucially, the central TQV core does not fragment: while its size oscillates in response to the global breathing motion, it persists as a single multiply quantized vortex line throughout the evolution. This stability holds even at relatively long times when transient density interference patterns become visible.

This behavior stands in stark contrast to the established instability of a TQV in a standard, non-ring-shaped (e.g., Thomas-Fermi) condensate, where it rapidly decays into three singly quantized vortices~\cite{Pethick:2002a}. The key stabilizing mechanism in our geometry is the confined, circulating superflow within the ring. The periodic inward and outward motion of the ring's density wall, driven by the initial toroidal preparation and the harmonic confinement, effectively suppresses the dynamical modes that would otherwise lead to vortex splitting. This stabilization of a multiply quantized vortex by a dynamically breathing ring-shaped density profile is consistent with our earlier findings for a doubly quantized vortex in a similar setup ~\cite{Xiong_2020}.

Therefore, the splitting dynamics analyzed in the main text is not an intrinsic instability of the TQV in our ring-shaped system but is selectively induced by breaking the rotational symmetry of the trap ($\eta>0$). The isotropic case ($\eta=0$) presented here establishes the baseline stable configuration, from which the controlled, anisotropy-driven instability explored in the main body proceeds.

\begin{figure*}[t]
  \centering
  \includegraphics[width=\textwidth]{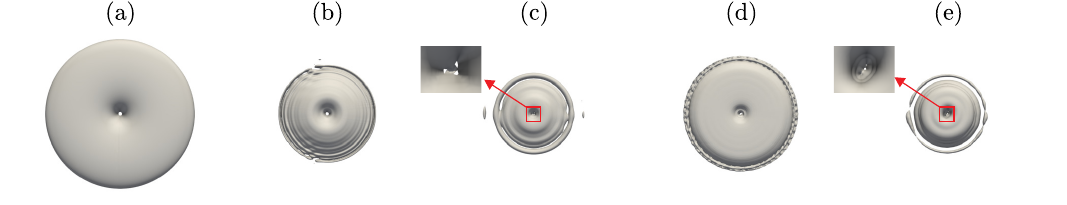}
  \caption{(Color online) Temporal density isosurface of a ring-shaped condensate with the circulation $S=3$ in a weakly asymmetric trap, i.e., $\eta=0.001$ for  $g=g_0$. Different panels correspond to different time : (a) $t=0~\mathrm{ms}$, (b) $t=0.186~\mathrm{ms}$, (c) $t=0.371~\mathrm{ms}$, (d) $t=1.855~\mathrm{ms}$, and(e) $t=2.65~\mathrm{ms}$.}
  \label{fig:process_of_tri}
\end{figure*}

\subsection*{Appendix B: Splitting dynamics of the TQV for $\eta = 0.001$}
\label{app:TQVV}

Fig.\,\ref{fig:process_of_tri} displays the time evolution of the top-view density isosurface for a ring-shaped condensate with circulation $S = 3$ in an extremely weak elliptical trap ($\eta = 0.001$). The condensate first undergoes a radial contraction, compressing the initial toroidal density profile [Figs.~\ref{fig:process_of_tri}(a) and (b)]. Driven by this contraction and the residual trap anisotropy, the TQV splits into three singly quantized vortices that initially adopt a triangular arrangement [Fig.~\ref{fig:process_of_tri}(c)]. This early-stage pattern reflects the dominant influence of the nearly isotropic geometry inherited from the initial ring-shaped preparation.

Subsequently, as the cloud expands and the central density decreases, the three vortices recombine into a single giant vortex [Fig.~\ref{fig:process_of_tri}(d)]. This recombination is transient. At later times, the persistent weak anisotropy of the trap progressively steers the instability, causing the vortex to split again---this time into a linear (chain-like) arrangement aligned with the soft axis of the trap [Fig.~\ref{fig:process_of_tri}(e)].

The observed sequence from transient triangular splitting, then recombination, to linear splitting illustrates how a minute breaking of rotational symmetry ($\eta = 0.001$) can qualitatively alter the dynamical pathway. The early triangular pattern is consistent with the near-isotropic $l_q=3$ Bogoliubov mode, while the final linear alignment emerges as the trap anisotropy selectively amplifies the $l_q=2$ channel on longer timescales.

\bibliography{Gdsd}

\end{document}